\documentclass{nature}
\usepackage{graphicx}% Include figure files
\usepackage{amssymb}
\usepackage{dcolumn}% Align table columns on decimal point
\usepackage{bm}% bold math
\usepackage{epstopdf}
\usepackage{epsfig}
\usepackage{color}
\usepackage[colorlinks=true, letterpaper=true, pdfstartview=FitV, linkcolor=blue, citecolor=blue, urlcolor=blue]{hyperref}
\newcommand{\onlinecite}[1]{\hspace{-1 ex} \nocite{#1}\citenum{#1}}

\begin{document}

\title{Anisotropic Quantum Confinement Effect and Electric Control of Surface States in Dirac
Semimetal Nanostructures}

\author{Xianbo Xiao$^1$, Shengyuan A. Yang$^{2,\star}$, Zhengfang Liu$^3$, Huili Li$^1$, Guanghui Zhou$^{4,\star}$}

\maketitle

\begin{affiliations}
\item School of Computer, Jiangxi University of Traditional Chinese Medicine, Nanchang 330004, China
\item Engineering Product Development, Singapore University of Technology and Design, Singapore 138682, Singapore
\item School of Basic Science, East China Jiaotong University, Nanchang 330013, China
\item Department of Physics and Key Laboratory for Low-Dimensional Quantum Structures and Manipulation (Ministry of Education), Hunan Normal University, Changsha 410081, China

$^\star$e-mail: shengyuan\_yang@sutd.edu.sg; ghzhou@hunnu.edu.cn
\end{affiliations}

\begin{abstract}
The recent discovery of Dirac semimetals represents a new achievement in our fundamental
understanding of topological states of matter. Due to their topological surface states,
high mobility, and exotic properties associated with bulk Dirac points, these new materials
have attracted significant attention and are believed to hold great promise for fabricating
novel topological devices. For nanoscale device applications, effects from finite size usually
play an important role. In this report, we theoretically investigate the electronic properties
of Dirac semimetal nanostructures. Quantum confinement generally opens a bulk band gap at the
Dirac points. We find that confinement along different directions shows strong anisotropic effects.
In particular, the gap due to confinement along vertical $c$-axis shows a periodic modulation,
which is absent for confinement along horizontal directions. We demonstrate that the topological
surface states could be controlled by lateral electrostatic gating. It is possible to
generate Rashba-like spin splitting for the surface states and to shift them relative to the
confinement-induced bulk gap. These results will not only facilitate our fundamental understanding
of Dirac semimetal nanostructures, but also provide useful guidance for designing all-electrical
topological spintronics devices.
\end{abstract}

In the past years, the discovery of novel topological states of matter has attracted
significant attention\cite{Hasan,Qi}. Topologically nontrivial states are first
demonstrated in the bulk insulating phases. The most salient feature of such topological
insulators (TIs) is that they have a full insulating gap in the bulk but gapless edge or
surface states at boundaries\cite{Hasan,Qi}. Therefore, the conductance of a system at Fermi energy is
supported only by the transport through the sample boundaries. With preserved time
reversal symmetry, the spectrum at the boundary typical shows linear energy dispersion,
and the electron spins are locked to their transport directions. For two-dimensional
(2D) TIs, the 1D edge hosts gapless spin-helical edge channels linearly crossing at a
Dirac point\cite{Konig}; whereas a 3D TI has protected spin-helical surface states usually with
a Dirac cone like dispersion\cite{Ando}. The Dirac character of the low energy quasiparticles is
underlying many intriguing physical properties of these topological materials. Recently,
such Dirac-type excitation has been extended to 3D, leading to the concept of topological
semimetals\cite{Murakami,Wan,Burkov}. These remarkable states have nontrivial topological properties without having
a bulk band gap. In particular, a class of Dirac semimetals (DSM) has proposed\cite{Rappe1}, which has
a Fermi surface consisting of four-fold degenerate 3D Dirac point(s) protected by the
crystalline symmetry. Several materials have been predicted to be DSMs, including
$\mathrm{A_{3}Bi}$ (A=Na, K, Rb) and $\mathrm{Cd_{3}As_{2}}$\cite{Wang,Wang2}. Experiments on
$\mathrm{Na_{3}Bi}$ \cite{Liu,Zhang2} and $\mathrm{Cd_{3}As_{2}}$ \cite{Borisenko,Neupane,Yi} have indeed successfully confirmed their
DSM character.

When the system size is comparable to the quasiparticle wavelength, quantum confinement
effect is going to be important. Since for quasiparticles near the Dirac point, the effective
wave number approaches zero, hence they should be strongly affected by quantum confinement.
Indeed, it has been shown that in finite size samples, the coupling between two edges
(or two surfaces) could generate gaps at the boundary Dirac points for 2D (or 3D) TIs\cite{Zhou,Linder,He,Lu,Liu3,Imura}.  One may
naturally expect that quantum confinement should also strongly affect DSMs. Moreover,
because now the Dirac point is in the bulk spectrum, the confinement would actually
transform the system to a bulk insulator, which is more drastic than that for TIs where
the change is mainly on the boundary.

In this work, motivated by the recent success in experimentally realizing these DSM
materials and by the progress in fabricating nanostructures with well-controlled sizes
and orientations, we investigate the properties of quantum confined DSM nanostructures.
We show that the confinement-induced gap has a strong aniostropic dependence on the
confining direction due to the intrinsic bulk anisotropy. Particularly,
the gap due to confinement along $c$-axis has a periodic modulation with the period
determined by the passing of subsequent quantum well states through the original Dirac
point. Further, we study the evolution of the boundary states and show that the edge channels
in this system is in fact advantageous over the usual 2D TIs in terms of robust gapless
edge transport. By using eletrostatic gating along transverse directions, it is possible
to generate Rashba-type spin-splitting in the edge states and also shift these states
relative to the bulk confinement-induced gap. The results presented here will be important
for utilizing these novel topological materials for nanoelectronics and spintronics applications.

\section*{Results}
\subsection{Low energy model and bulk spectrum.}

\begin{figure}
  \begin{center}
   \epsfig{file=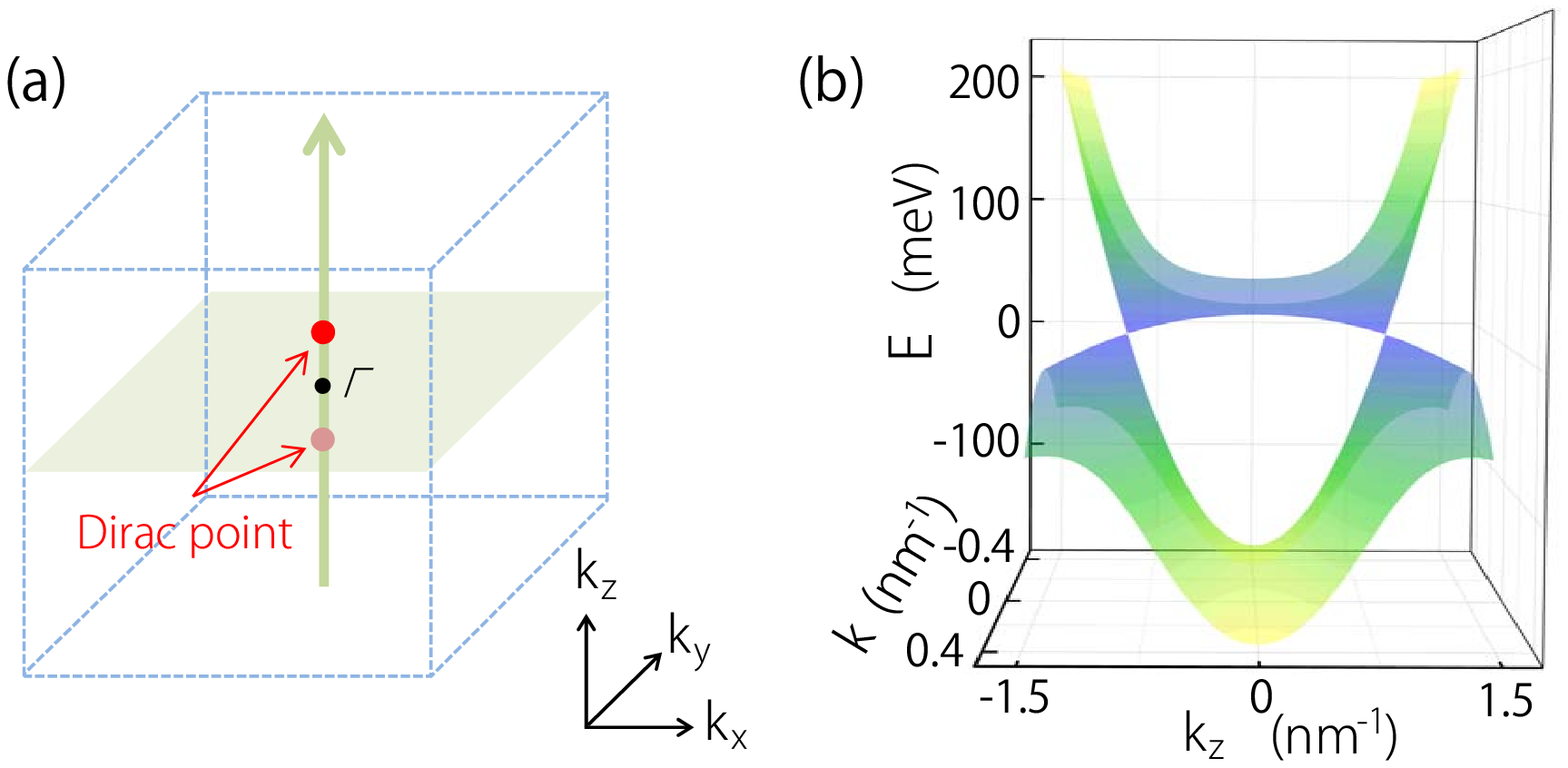,width=14cm}
  \end{center}
  \caption{\label{Fig1} (a) Schematic figure showing the locations of
  two Dirac points in Brillouin zone. The two points are located along $k_z$-axis (at $\bm k=(0,0,\pm k_\mathrm{D})$) and are symmetric about $\Gamma$-point.
  (b) Energy spectrum plotted as a function of
  $k_z$ and the wave vector $k$ (the wave vector in $k_x$-$k_y$ plane), obtained from the low energy model Eq.(\ref{H0}). The parameters for Na$_3$Bi are used in the calculation, as described in the main text.}
\end{figure}

We start from the low energy effective model of DSM materials derived from first
principles results for $\mathrm{A_{3}Bi}$ (A=Na, K, Rb) and $\mathrm{Cd_{3}As_{2}}$\cite{Wang,Wang2}.
The low energy states near the Fermi level consist of the four orbital basis
$|S_{\frac{1}{2}},\frac{1}{2}\rangle$, $|P_{\frac{3}{2}},\frac{3}{2}\rangle$,
$|S_{\frac{1}{2}},-\frac{1}{2}\rangle$ and $|P_{\frac{3}{2}}, -\frac{3}{2}\rangle$.
Around $\Gamma$ point in the Brillouin zone, the effective Hamiltonian expanded up
to quadratic order in $k$ is given by
\begin{equation}\label{H0}
H(\bm k)=\epsilon_0(\bm{k})+\left[\begin{array}{cccc}M(\bm{k}) & Ak_{+} & 0  & 0 \cr
                                                              Ak_{-} & -M(\bm{k}) & 0 & 0  \cr
                                                              0 & 0 & M(\bm{k}) & -Ak_{-}  \cr
                                                              0 & 0 & -Ak_{+} & -M(\bm{k}) \\
                                                              \end{array}\right],
\end{equation}
where
$\epsilon_{0}(\bm{k})=C_{0}+C_{1}k_{z}^{2}+C_{2}(k_{x}^{2}+k_{y}^{2})$,
$k_{\pm}=k_{x}\pm ik_{y}$, and $M(\bm{k})=M_{0}-M_{1}k_{z}^{2}-M_{2}(k_{x}^{2}+k_{y}^{2})$
with parameters $M_{0}, M_{1}, M_{2}<0$ to reproduce the band inversion feature. The parameters
$A$, $C_i$ and $M_i$ can be obtained by fitting the $\emph{ab initio}$ result for a specific
material. Energy spectrum of this Hamiltonian can be readily solved and is given by
$
E(\bm{k})=\epsilon_{0}(\bm{k})\pm\sqrt{M(\bm{k})^{2}+A^{2}k_{+}k_{-}}.
$
Two Dirac points can be obtained by requiring the square root term on the right hand side
to vanish, and they are located at $\pm\bm{k}_\mathrm{D}=(0, 0, \pm k_\mathrm{D})$ with
$k_\mathrm{D}=\sqrt{M_0/M_1}$. The two Dirac points are on the $k_z$-axis, symmetric about
the $\Gamma$ point (see Fig.~\ref{Fig1}(a)). Each of them is four-fold degenerate and can
be regarded as two overlapping Weyl points with opposite chiralities\cite{Rappe1} (represented by the
two $2\times 2$ diagonal blocks in Eq.(\ref{H0})). One observes that the spectrum is
isotropic in $k_x$-$k_y$ plane but is anisotropic for out-of-plane directions. In
Fig.~\ref{Fig1}(b), we plot the bulk energy spectrum as functions of $k_z$ and wave
vector $k=\sqrt{k_x^2+k_y^2}$ in $k_x$-$k_y$ plane. One can clearly observe the two
Dirac points along $k_z$-axis around which the dispersion is linear. Both the distribution
of the Dirac points and the energy dispersion are highly anisotropic. Here and in the
following calculations, we take the model parameters for $\mathrm{Na_{3}Bi}$ obtained
from the first-principles calculations,\cite{Wang} namely $C_{0}=-63.82$ meV, $C_{1}=87.536$
meV$\cdot\mathrm{nm^{2}}$, $C_{2}=-84.008$ meV$\cdot\mathrm{nm^{2}}$, $M_{0}=-86.86$ meV,
$M_{1}=-106.424$ meV$\cdot\mathrm{nm^{2}}$, $M_{2}=-103.610$ meV$\cdot\mathrm{nm^{2}}$,
and $A=245.98$ meV$\cdot\mathrm{nm}$. This leads to $k_\mathrm{D}=0.903$ nm$^{-1}$. These parameters agree well with the recent experimental result\cite{Liu}. We also tested the parameters for another DSM Cd$_3$As$_2$, which shares qualitatively similar low energy physics as $\mathrm{Na_{3}Bi}$.
The essential features discussed below applies to Cd$_3$As$_2$ as well.

\subsection{Confinement along horizontal direction.}

\begin{figure}
  \begin{center}
   \epsfig{file=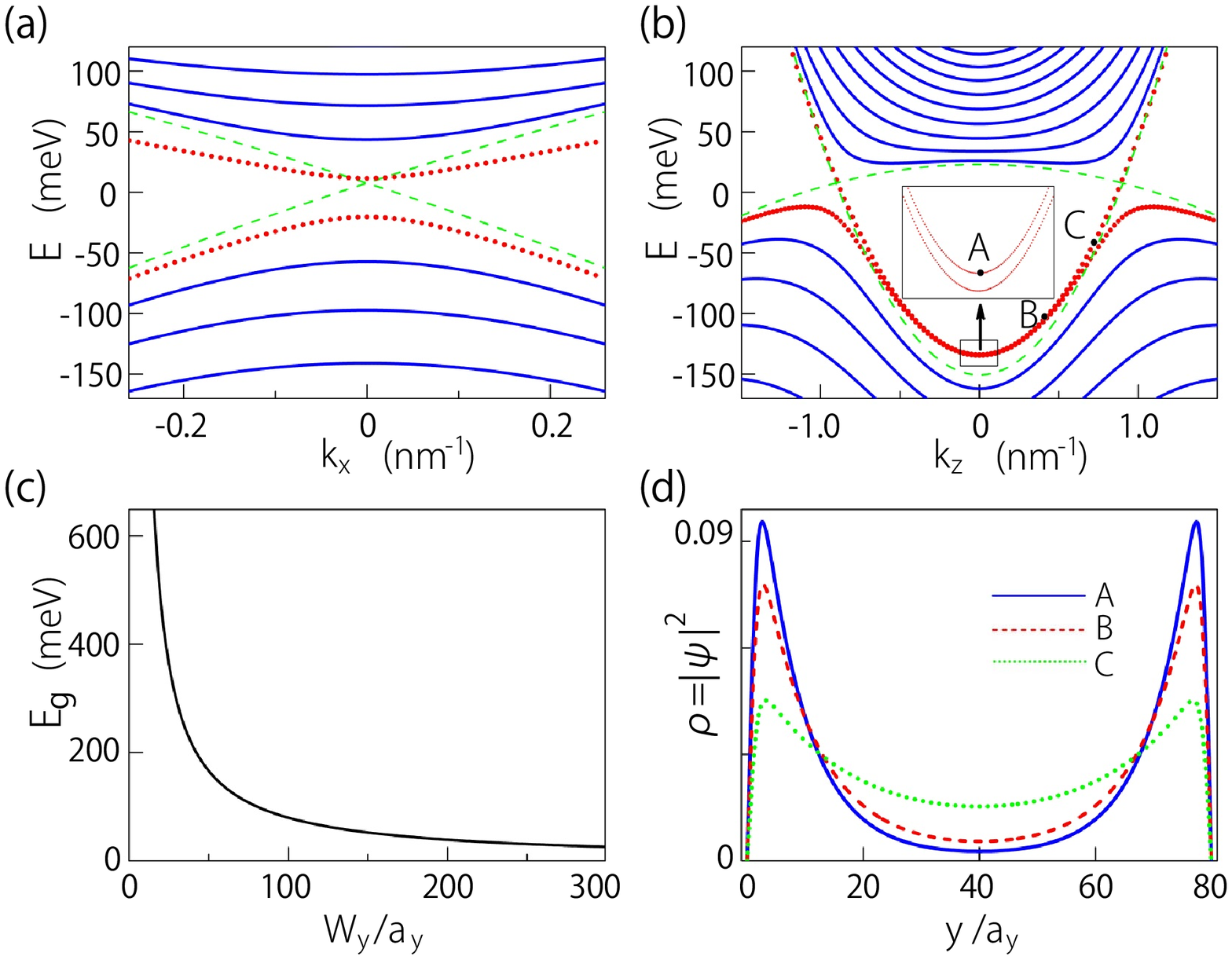,width=14cm}
  \end{center}
  \caption{\label{Fig2} The case of confinement along $y$-direction. (a) Energy spectrum along
  $k_x$-direction through the original Dirac point $(0,0,k_\mathrm{D})$. (b) Energy spectrum along $k_z$-direction
  at $k_x=0$.
  In (a) and (b), $W_y=80a_y$, the green dashed curves are the original bulk bands, the blue curves are the confinement-induced subbands, and the red
  dotted curves are the surface bands. Inset in (b) shows the small gap at $\Gamma$ point due to hybridization between two opposite surfaces.
  A, B, and C label the states on the upper surface band. (c) Confinement-induced bulk gap versus the
  confinement width $W_y$. (d) Electron density distribution along the confinement direction corresponding to the states A, B, and C as indicated in (b).}
\end{figure}

To investigate the quantum confinement effect, we discretize the Hamiltonian on a 3D
simple cubic lattice. We take the lattice constants as $a_{x}=a_{y}=0.3$ nm and
$a_{z}=0.5$ nm. We first consider the case where the system is confined along one
direction. Since the model is isotropic in $k_x$-$k_y$ plane, in Fig.~\ref{Fig2},
we choose a finite width $W_y$ along $y$-direction and calculate the system spectrum.
Due to the confinement, the continuum spectrum now splits into subbands. This
opens up a finite band gap for the bulk at the Dirac points. In Fig.~\ref{Fig2}(a)
and \ref{Fig2}(b), we show the energy spectra crossing one Dirac point $\bm k_\mathrm{D}$
along $k_x$ and $k_z$ directions respectively. The confinement-induced gap can
be clearly observed. The green dashed curves indicate the original bulk bands.
One also notice that there exist gapless topological surface states due to the band
inversion feature of Hamiltonian (\ref{H0}). These states are marked as red dotted
curves. The probability density distribution of three selective states on the surface
band is plotted in Fig.~\ref{Fig2}(d). One observes that these states are indeed
peaked around the two surfaces $y=0$ and $y=W_y$. Note that because the system has
both time reversal and inversion symmetries, hence each point in fact has two
degenerate states localized on each surface, and in Fig.~\ref{Fig2}(d), we choose
to plot their symmetric superpositions. The localization of the surface states
decreases from $k_z=0$ with increasing $k_z$ values, as can be seen by comparing
the density distribution at different $k$ points. Similar to that for the 2D TIs\cite{Zhou},
due to the finite width, the states on two surfaces have a small hybridization,
which opens a small gap at $\Gamma$ point, as shown in the inset of Fig.~\ref{Fig2}(b).
Moreover, one can see that the (avoided) Dirac point for the surface states are
buried in the valence band. As a result, there are always surface states crossing
the bulk confinement-induced gap. This in fact makes the system similar to the usual
3D TI such as Bi$_2$Te$_3$\cite{Zhang}. The surface bands in the gap are protected by time
reversal symmetry from opening a gap.

The induced bulk gap depends on the confinement width $W_y$. Using the quantum well
approximation, the effective wave vector for the first few quantum well modes have
$k_y\simeq n\pi/W_y$ with $n=1,2,\cdots$. Because the Dirac points are located
along $k_z$-axis and the dispersion around them are linear, the induced gap should
roughly scale with $W_y$ as $E_g\sim W_y^{-1}$. This feature is indeed confirmed as
in Fig.~\ref{Fig2}(c), where we plot the bulk confinement-induced gap as a function
of $W_y$. The gap shows a monotonic decay with increasing $W_y$.

\subsection{Confinement along vertical direction.}

\begin{figure}
  \begin{center}
   \epsfig{file=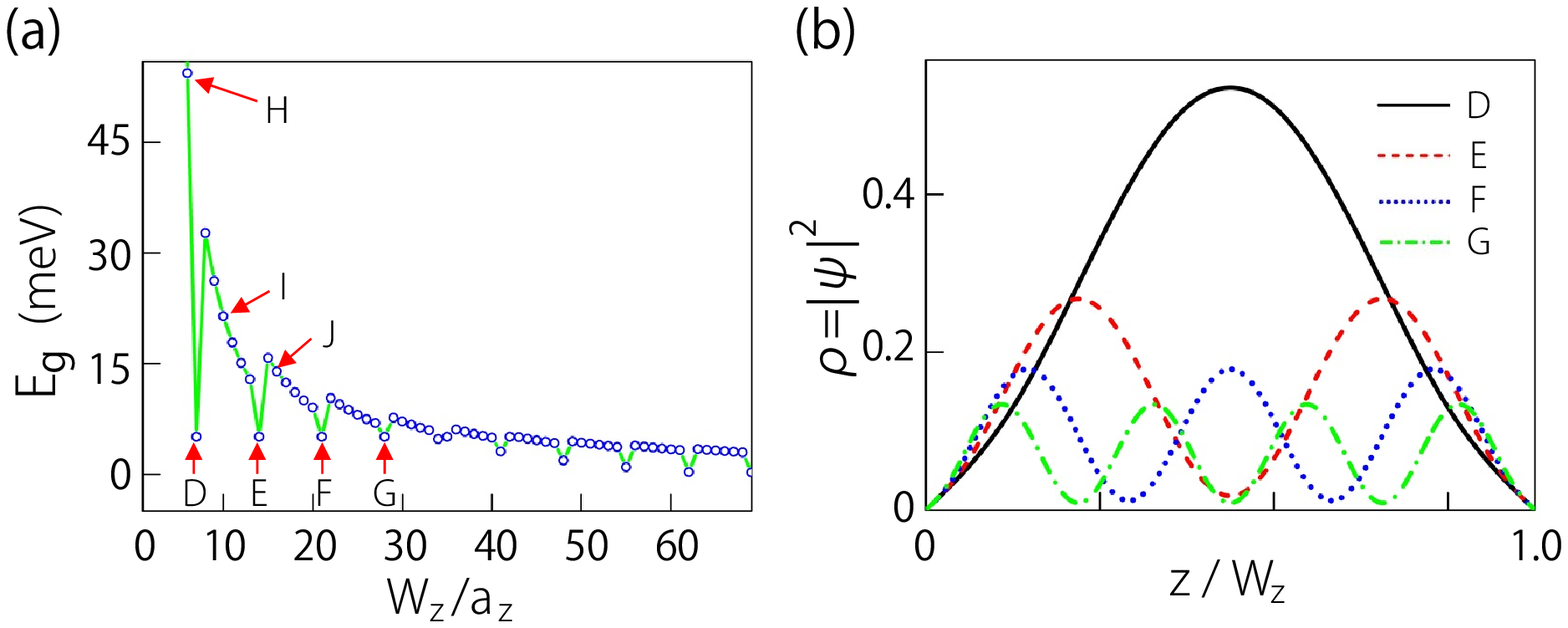,width=14cm}
  \end{center}
  \caption{\label{Fig3} The case of confinement along $z$-direction. (a) Confinement-induced
  bulk gap versus confinement thickness $W_z$, showing a periodic modulation. (b) Electron density
  distribution along the confinement direction of states at the valence band edge corresponding to the systems with different thickness D, E, F, and G as indicated in (a).}
\end{figure}

Next, we consider the system with only confinement along $z$-direction. In this case,
because the two Dirac points projects to the same point on the surface, there is
no topological surface state on the $z=0$ and $z=W_z$ sufaces. The bulk subbands is
isotropic in the $k_x$-$k_y$ plane, resembling that in Fig.~\ref{Fig2}(a), and the
situation seems quite trivial. However, if we again plot the confinement induced gap
versus the confinement width $W_z$, we observe that in contrast to the case in
Fig.~\ref{Fig2}(c), now the gap has a periodic modulation with a period of $7a_z$
superposed upon the power law decay (see Fig.~\ref{Fig3}(a)). To understand this
interesting feature, we note that the two Dirac points are located at finite distance
$k_\mathrm{D}$ from the $\Gamma$ point along $k_z$-axis. The minimum of the gap occurs
around the Dirac points. With increasing width $W_z$, the wave vector of each subband
is decreasing and is going to sweep across the Dirac point, leading to a dip on the
curve of $E_g$. Using quantum well approximation for the effective wave vector of
subbands $k_z\simeq n\pi/W_z$, one can easily find that the period of the oscillation
should be $\pi/(k_\mathrm{D}a_z)\approx 7$, which nicely explains the observation in
Fig.~\ref{Fig3}(a). To further confirm this physical picture, we check the density
distribution of the band edge states corresponding to the dips labeled in Fig.~\ref{Fig3}(a).
As shown in Fig.~\ref{Fig3}(b), we see that for a sequence of dips, the corresponding
density distributions show increasing number of nodes along the confinement direction,
indicating that the states are from subbands with increasing quantum well mode number
$n$, which is consistent with our analysis.

Because band inversion occurs for $k_z$ between the two Dirac points, the band gap for
a subband is inverted if its $k_z$ is less than $k_\mathrm{D}$. If we consider the system
confined along $z$-direction as a quasi-2D system, then the periodic modulation we
discussed before is also accompanied with an oscillation in its $Z_2$ topological
invariant\cite{Wang2}. With increasing $W_z$, for each quantum well subband with $k_z$ crossing
$k_\mathrm{D}$, it adds a $Z_2=1$ to the total 2D $Z_2$ number, leading to the oscillation
its topological property. Specifically, from Fig.~\ref{Fig3}(a), we mentioned that the
$n=1$ quantum well subband passes $k_\mathrm{D}$ around the thickness at point D, the $n=2$
subband passes $k_\mathrm{D}$ around point E, and so on. For the thickness at point H, no
subband has band inversion, hence the 2D $Z_2$ invariant must be zero. For thickness
between D and E, one subband becomes topological nontrivial, hence $Z_2=1$.  For
thickness between E and F, one more subband becomes nontrivial, leading to a total
$Z_2=0$, i.e. the system again becomes trivial. The oscillatory behavior in topological character for Na$_3$Bi was briefly mentioned in Ref.~\onlinecite{Wang}, and a more detailed discussion was presented for Cd$_3$As$_2$ in Ref.~\onlinecite{Wang2}.
Our results are consistent with these previous studies.

\begin{figure}
  \begin{center}
   \epsfig{file=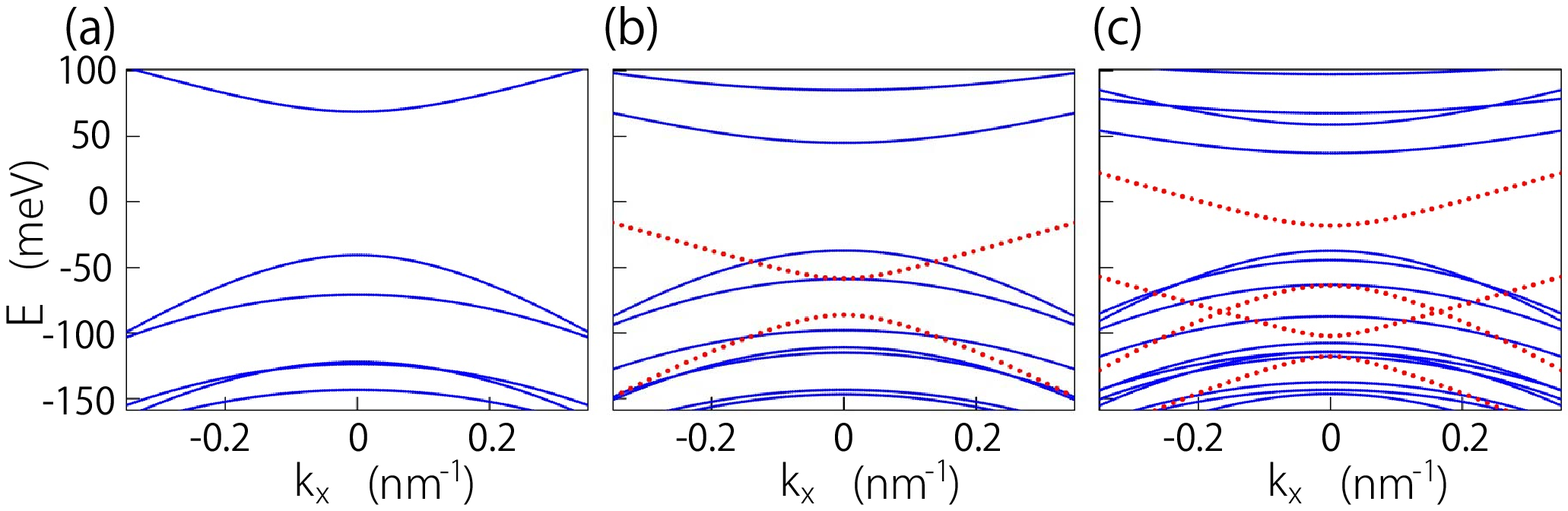,width=14cm}
  \end{center}
  \caption{\label{Fig4} Energy spectrum of the system confined in both $y$ and $z$
  directions, with $W_y=40a_y$ and $W_z$ given by the thickness corresponding to point (a) H, (b) I, and (c) J as in
  Fig.~\ref{Fig3}(a). The blue curves are the bulk subbands and the red dotted curves denote the edge states.}
\end{figure}

The oscillation in $Z_2$ invariant can be better visualized by considering the system
confined in both $y$ and $z$ directions. In Fig.~\ref{Fig4}, we plot the spectra for
the three thickness $W_z$ labeled by H, I, and J in Fig.~\ref{Fig3}(a), with a width
$W_y=40a_y$. For point H, since all the subbands are trivial, there is no edge states
on the side surfaces $y=0$ and $y=W_y$. For point I, one subband becomes nontrivial,
as a result, there is a pair of spin-helical edge states on the side surfaces, as
indicated by the red dotted curves in Fig.~\ref{Fig4}(b). One notes that there is a
small gap opened at $k_x=0$ for the edge bands, which is due to the hybridization
between the states on the two side surfaces. Note that because the (avoided) Dirac point of the edge bands
is buried in the valance bands (at least for Na$_3$Bi and Cd$_3$As$_2$). There are always gapless
topological edge states traversing the confinement-induced bulk gap. These states are protected by time reversal symmetry
and forming dissipationless edge transport channels. For point J, two
subbands have band inversion, each resulting in a pair of helical edge states.
Hence in the spectrum, we observe that there are two pairs of helical edge states.
Because now scattering between states from different pairs are allowed by symmetry,
these edge states lose the topological protection hence are not as robust as that
in Fig.~\ref{Fig4}(b).

\subsection{Electric control of boundary channels.}
We have analyzed the confinement induced bulk gap, and also show that robust
topological boundary channels can exist inside the bulk gap. These channels
could be utilized for designing topological devices. For device applications,
it is desirable to have full electric control because electric control generally
has fast response, consumes less energy, and is more easily to be implemented
and integrated. In the following, we investigate the possibility to control the
boundary states of confined DSM through electrostatic means.

\begin{figure}
  \begin{center}
   \epsfig{file=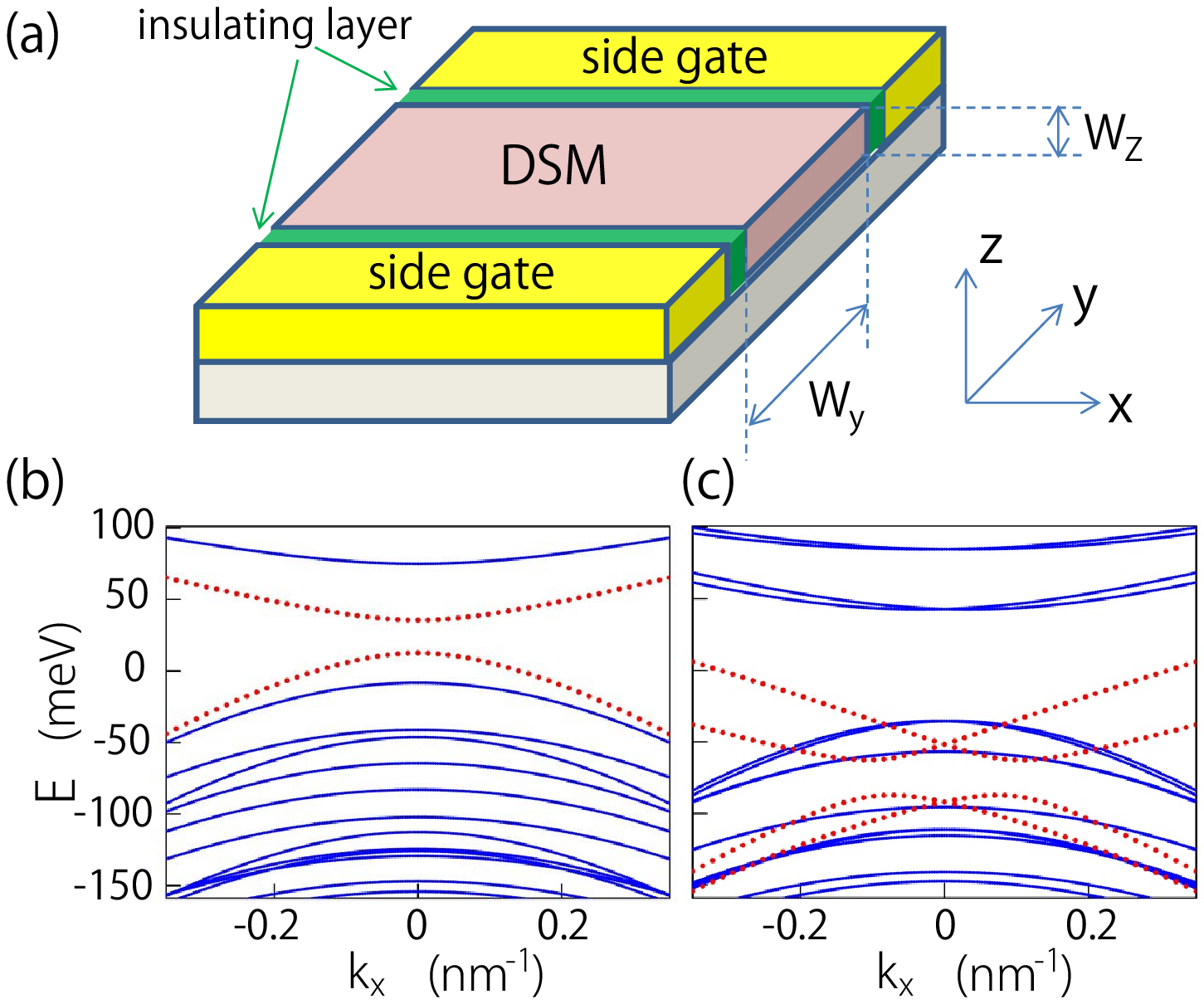,width=14cm}
  \end{center}
  \caption{\label{Fig5} (a) Schematic figure showing a possible
  transverse gating setup. The DSM nanostructure has a width of $W_y$ and a thickness of $W_z$.
  The structure is assumed to be extended along $x$-direction which is the
  transport direction. (b) Energy spectrum of the model with surface potential $2.1$ V
  applied on the two side-surfaces. (c) Energy spectrum of the model when a transverse
  electric field with $E=3.75\times 10^{-2}$ V/$a_y$ is applied. Here $W_y=40a_y$,
  $W_z=10a_z$. The blue curves are the bulk subbands and the red dotted curves denote the edge states.}
\end{figure}

We first discuss the case by using electric gating on the side surface to tune
the surface potential. Let's consider a nanostructure confined in both $y$ and
$z$ directions with a nontrivial 2D $Z_2$ invariant as that in Fig.~\ref{Fig4}(b).
We model the surface potential by gating by applying an equal onsite potential
in the lattice model for the two surface layers at $y=0$ and $y=W_y$. Because
the boundary states are localized near the two side surfaces, they are most
susceptible to the applied surface potential than the bulk states. As a result,
we could tune the position of the boundary states relative to the confined-induced
bulk gap. In Fig.~\ref{Fig5}(a), a surface potential of $2.1$ V is applied, and
we see that the (avoided) edge band Dirac point is shifted into the bulk gap.
If the Fermi level is inside both the bulk gap and the surface gap, the transport
through the system could be turned off. This means that we could turn the edge
transport on and off electrically, which is much desired for application purpose.

When the two gates are at different potential, a transverse electric field
along $y$-direction could be applied to the system. This obviously breaks the
inversion symmetry. Therefore the double degeneracy is going to be lifted.
Here we model the transverse field as a simple linearly varying onsite potential
term $V(y)= E\cdot(y-W_y/2)$. As shown in Fig.~\ref{Fig5}(b), both the bulk
subbands and the boundary states have a Rashba type spin splitting induced
by the electric field. We have analyzed the spin polarizations of these states
and the spin texture indeed coincides with that for Rashba spin-splitting. This
result is analogous to the situation discussed in the
context of 2D TIs\cite{Liu2}. These spin-polarized edge channels will be useful for
spintronics applications. We mention that the modeling of electrostatic effects
above is based on a very simple modeling. For more realistic analysis, one needs
to solve the electronic structure combined with Poisson's equation self-consistently.
Nevertheless, the essential features we discussed above should still apply.

\section*{Discussion}
We have shown that the confinement effect generally opens up a gap at the Dirac
points, which would be desired for the purpose of designing logical devices. In
previous studies of 2D and 3D topological insulators, one aim is to realize robust
gapless topological edge or surface states in the bulk gap, which can be used for
energy-efficient charge or spin transport. However, it has been found that due to
confinement, the states on the opposite edge or surfaces can hybridize and this
opens up a gap for the boundary states\cite{Zhou,Linder}. In typical 2D TIs such as HgTe/CdTe quantum wells, the edge state crossing occurs in the bulk gap. Hence when a gap is opened by coupling between the two edges, the edge channels would no longer be gapless.
This was considered as a detrimental effect for applications\cite{Zhou,Liu2}. In this regard, the DSM nanostructures is
in fact advantageous in that its edge state crossing is buried in the valence band hence the edge transport inside bulk gap would not be affected (at least for the cases of Na$_3$Bi and Cd$_3$As$_2$). As a result, there
are always gapless topological edge channels in the confinement-induced bulk gap which are protected by time reversal symmetry.
This implies that the DSM nanostructures could offer a better platform for designing
topological electronic devices with low power dissipation. Note that the above comparison is made in the native states of the systems. We also mention that the detailed edge states dispersions would depend on the edge/surface condition. For example, the edge sates could be shifted in energy by electric gating as we shown in the previous discussion, and be distorted by surface defects or doping\cite{ZKLiu}.

To conclude, we have investigated the electronic properties of confined DSM
nanostructures. We find that the quantum confinement effect is strongly anisotropic.
The confinement generally induce a bulk gap at the Dirac points and it shows a
periodic modulation when the confinement is along $c$-axis. This modulation is
accompanied with topological transitions. For a topological nontrivial state,
we point out that the system is in fact superior to usual 2D TIs in terms of robust
gapless edge transport. Electric means to manipulate the edge channels are discussed.
Our result show that DSM nanostructures offer a promising platform for designing novel
topological quantum devices.

\section*{Methods}
\subsection{Lattice model.}
To investigate the confinement effects, we discretize the low energy continuum model in Eq.(\ref{H0}) on a 3D tetragonal lattice
with lattice constants $a_x$, $a_y$ and $a_z$ along the three orthogonal directions.
\begin{eqnarray}
H_\mathrm{lattice}=\sum\limits_{i,j,k}V_{i,j,k}c_{i,j,k}^{\dag}c_{i,j,k}+\sum\limits_{i,j,k}
(T_{x}c_{i+1,j,k}^{\dag}c_{i,j,k}+T_{y}c_{i,j+1,k}^{\dag}c_{i,j,k}+T_{z}c_{i,j,k+1}^{\dag}c_{i,j,k}+\mathrm{H.c.}),
\end{eqnarray}
where $c_{i,j,k}^{\dag}(c_{i,j,k})$ is the creation (annihilation) operator for an electron on the site
$(i,j,k)$. The on-site matrix element $V_{i,j,k}$ is given by
\begin{eqnarray}
V_{i,j,k}=
\left[\matrix{E_{+} & 0 & 0 & 0 \cr
             0 & E_{-} & 0 & 0 \cr
             0 & 0 & E_{+} & 0 \cr
             0 & 0 & 0 & E_{-} }\right],
\end{eqnarray}
where $E_{\pm}=C_{0}\pm M_{0}+2(C_{2}\mp M_{2})/a_{x}^{2}+2(C_{2}\mp M_{2})/a_{y}^{2}+2(C_{1}\mp M_{1})/a_{z}^{2}$. The $C$'s and $M$'s (and the $A$ below) are the material specific parameters defined in the main text.
The nearest-neighbor hopping matrix elements $T_{x}$, $T_{y}$ and $T_{z}$ are given respectively as
\begin{eqnarray}
T_{x}=
\left[\matrix{(-C_{2}+M_{2})/a_{x}^{2} & -iA/(2a_{x)}) & 0 & 0 \cr
             -iA/(2a_{x}) & (-C_{2}-M_{2})/a_{x}^{2} & 0 & 0 \cr
             0 & 0 & (-C_{2}+M_{2})/a_{x}^{2} & iA/(2a_{x)}) \cr
             0 & 0 & iA/(2a_{x)}) & (-C_{2}-M_{2})/a_{x}^{2} }\right],
\end{eqnarray}
\begin{eqnarray}
T_{y}=
\left[\matrix{(-C_{2}+M_{2})/a_{y}^{2} & A/(2a_{y)}) & 0 & 0 \cr
             -A/(2a_{y}) & (-C_{2}-M_{2})/a_{y}^{2} & 0 & 0 \cr
             0 & 0 & (-C_{2}+M_{2})/a_{y}^{2} & A/(2a_{y)}) \cr
             0 & 0 & -A/(2a_{y)}) & (-C_{2}-M_{2})/a_{y}^{2} }\right]
\end{eqnarray}
and
\begin{eqnarray}
T_{z}=
\left[\matrix{(-C_{1}+M_{1})/a_{z}^{2} & 0 & 0 & 0 \cr
             0 & (-C_{1}-M_{1})/a_{z}^{2} & 0 & 0 \cr
             0 & 0 & (-C_{1}+M_{1})/a_{z}^{2} & 0 \cr
             0 & 0 & 0 & (-C_{1}-M_{1})/a_{z}^{2} }\right].
\end{eqnarray}
The electrostatic effects are modeled by adding an on-site electrostatic potential energy $V_{i,j,k}$ $\rightarrow$ $V_{i,j,k}+V_{i,j,k}^\mathrm{E}$.
Using this lattice Hamiltonian, the low energy spectrum and the wavefunctions
of electronic states for the DSM nanostructures can be obtained conveniently.

\bibliographystyle{apsrev4-1}

\begin{addendum}
\item [Acknowledgments]
This work was supported by NSFC (Grant Nos. 11264019, 11364019, 11464011 and 11274108), by the
development project on the young and middle-aged teachers in the
colleges and universities in Jiangxi, and by SUTD-SRG-EPD2013062.

\item [Author Contributions]
X.X. and S.A.Y. conceived the idea. X.X. performed the calculation and the data analysis.
X. X., S.A.Y., and G.H.Z. contributed to the interpretation of the results and
wrote the manuscript. Z.F.L. and H.L.L. contributed in the discussion.
All authors reviewed the manuscript.

\item [Competing Interests]
The authors declare no competing financial interests.

\item [Correspondence]
Correspondence and requests for materials should be addressed to Shengyuan A. Yang or Guanghui Zhou.

\end{addendum}

\clearpage
\newpage

\newpage

\newpage

\newpage

\newpage

\end{document}